\documentclass[aps,pra,twocolumn,eqsecnum,showpacs,superscriptaddress,floatfix,nofootinbib]{revtex4}

\usepackage{psfrag,graphicx, rotating}
\usepackage{dcolumn}
\usepackage{amsmath,amssymb}
\usepackage{bm, float}
\usepackage[dvips]{epsfig,color}
\usepackage[english]{babel}

\newcommand{\be}{\begin{equation}}
\newcommand{\bea}{\begin{eqnarray}}
\newcommand{\eea}{\end{eqnarray}}
\newcommand{\ee}{\end{equation}}

\begin{document}

\title{Entanglement and majorization in (1+1)-dimensional quantum systems}

\author{Rom\'an Or\'us} 
\affiliation{Dept. d'Estructura i Constituents de la Mat\`eria, Univ. Barcelona, 08028, Barcelona, Spain.}  

\date{\today}

\begin{abstract}

Motivated by the idea of entanglement loss along Renormalization Group flows,
analytical majorization relations are proven for the ground state of
$(1+1)$-dimensional conformal field theories. For any of these theories, 
majorization is proven to hold in the spectrum of the reduced density matrices 
in a bipartite system when changing the size $L$ of one of the subsystems. 
Continuous majorization along uniparametric flows is also proven as long 
as part of the conformal structure is preserved under the deformation and some 
monotonicity conditions hold as well. As particular examples of our derivations, 
we study the cases of the \emph{XX}, Heisenberg and \emph{XY} 
quantum spin chains. Our results provide in a rigorous way
explicit proves for all the majorization 
conjectures raised by Latorre, L$\ddot{{\rm u}}$tken, Rico, Vidal and 
Kitaev in previous papers on quantum spin chains. 

\end{abstract}


\maketitle

\section{Introduction}

In the last few years, the emerging field of quantum information science 
\cite{chuang} has developed tools and
techniques for the analysis of quantum systems which have been proved to be
useful in other fields of physics. The study of many-body
Hamiltonians, quantum phase transitions, and the quantum correlations
-or entanglement- that these systems develop, are examples of this
interdisciplinary research. In fact, understanding entanglement has been
realized as one of the most 
challenging and interesting problems in physics \cite{Pres}. 

Another interesting application of the tools of quantum
information science has been the use of majorization theory \cite{maj} in order to
analyze the structure present in the ground state -also called vacuum- of some models along
Renormalization Group (RG) flows (for a recent review on RG, see
\cite{carteret}). Following this idea, Latorre et
al. \cite{fine-grained} proposed that irreversibility along RG
flows may be rooted in properties of the vacuum only, without necessity of
accessing the whole Hamiltonian of the system and its excited states. The vacuum of a theory
may already have enough information in order to envisage irreversibility along RG
trajectories. Such an irreversibility was casted into the idea of an \emph{
  entanglement loss} along RG flows, which proceeded in three constructive steps for
(1+1)-dimensional quantum systems: first, due to the fact that the central
charge of a (1+1)-dimensional conformal field theory is in fact a genuine
measure of the bipartite entanglement present in the ground state of the
system \cite{VLRK02,korepin-cardy}, there is a global loss of
entanglement due to the c-theorem of Zamolodchikov \cite{ctheorem}, which
assures that the value of the central charge at the ultraviolet fixed point is bigger or
equal than its value at the infrared fixed point ($c_{UV} \ge
c_{IR}$); second, given a splitting of the system into two contiguous pieces, there is a 
monotonic loss of entanglement due to the
monotonicity numerically observed for the entanglement entropy between the
two subsystems along the
flow, decreasing when going away from the critical fixed (ultraviolet) point;
third, this loss of entanglement is seen to be fine-grained, since it follows
from a very strict set of majorization ordering relations, which the
eigenvalues of the reduced density matrix of the subsystems are numerically seen to perfectly
obey. This last step motivated the authors of \cite{fine-grained} to affirm
that there was a \emph{ fine-grained entanglement loss} along RG flows rooted
in properties of the vacuum, at least for (1+1)-dimensional quantum systems.  
In fact, a similar fine-grained entanglement loss had already been numerically
observed by Vidal et al. in \cite{VLRK02}, for changes in 
the size of the bipartition described by the corresponding ground-state
density operators, at conformally-invariant critical points.

In this work, we analytically prove the links between conformal field theory (CFT), 
RG and entanglement that were conjectured in the recent papers \cite{VLRK02,fine-grained} 
for quantum spin chains.
We develop, in the bipartite scenario, a detailed and analytical study 
of the majorization properties of the eigenvalue spectrum obtained 
from the reduced density matrices of the ground state for a variety of 
$(1+1)$-dimensional quantum models in the bulk. 
Our approach is based on infinitesimal variations of the parameters defining
the model -magnetic fields, anisotropies- or deformations in the size of the block 
$L$ for one of the subsystems. We prove in these situations that there 
are strict majorization relations
underlying the structure of the eigenvalues of the considered reduced density 
matrices or, as defined in \cite{fine-grained}, there is a fine-grained entanglement loss. 
The result of our study is presented in terms of two
theorems. On the one hand, we prove exact continuous majorization
relations in terms of deformations of the size of the block $L$ that is
considered. On the other hand, we are also able to prove continuous 
majorization relations as a function of the parameters defining the model.
On top we also provide explicit analytical examples for models with a boundary based on 
previous work of Peschel, Kaulke and Legeza \cite{peschel1}.

This paper is structured as follows: in sec.II we remember the concepts of
global, monotonous and fine-grained entanglement loss, as defined in
\cite{fine-grained}. In sec.III we analytically prove
continuous majorization relations for any $(1+1)$-dimensional CFT 
when the size of the subsystem $L$ is changed, and give an example of a
similar situation for the case of the \emph{XX}-model with a boundary. In sec.IV we prove
continuous majorization relations with respect to the flows in parameter
space for $(1+1)$-dimensional quantum systems under perturbations which preserve 
part of the conformal structure of the partition function. Again, we support
our result with the analysis of a similar situation for the Heisenberg and 
\emph{XY} quantum spin chains with a boundary. 
Finally, sec.V collects the conclusions of our study. We also
review in appendix A the definition of majorization and provide 
a lemma which will be used in our calculations.

\section{Global, monotonous and fine-grained entanglement loss}

Consider the pure ground state (or vacuum) $|\Omega \rangle$ of a given 
physical system which depends on a particular set of parameters, and let us
perform a bipartition of the system into two pieces $A$ and $B$. The density
matrix for $A$, describing all the physical observables
accessible to $A$, is given by $\rho_A = {\rm tr}_B (|\Omega \rangle
\langle \Omega|)$ -and analogously for $B$-. In this section we will focus our
discussion on the density matrix for the subsystem $A$, so we will drop the
subindex $A$ from our notation. Let us consider a change in
one -for simplicity- of the parameters on which the resultant 
density matrix depends, say, parameter ``$t$'', which can be either 
an original parameter of the system or the size of the region $A$. In other words, we make the change
$\rho(t_1) \rightarrow \rho(t_2)$, where $t_1 \ne t_2$. In order to simplify
even more our discussion let us assume that $t_2 > t_1$. We wish to understand how this
variation of the parameter alters the inner structure of the ground state and,
in particular, how does it modify the entanglement between the two parties $A$
and $B$. Because we are considering entanglement at two different points $t_2$
and $t_1$, we assume for simplicity that the entanglement between $A$ and $B$ is
bigger at the point $t_1$ than at the point $t_2$, so we have an entanglement
loss when going from $t_1$ to $t_2$. 

Our characterization of this entanglement loss will progress through three
stages, as in \cite{fine-grained}, refining at every step the underlying 
ordering of quantum correlations. These
three stages will be respectively called \emph{global}, \emph{monotonous}
and \emph{fine-grained} entanglement loss. 

\bigskip

\paragraph{Global entanglement loss.-} The simplest way to quantify the loss
of entanglement between $A$ and $B$ when going from $t_1$ to $t_2$ is by means
of the entanglement entropy $S(\rho (t)) = -{\rm tr}(\rho (t) \ln{\rho (t)})$.
Since at $t_2$ the two parties are less entangled than at $t_1$, we have that 

\begin{equation}
S(\rho(t_1)) > S(\rho(t_2)) \ ,
\label{global}
\end{equation}
which is a global assessment between points $t_2$ and $t_1$. This is what we
shall call global entanglement loss. 

\bigskip

\paragraph{Monotonous entanglement loss.-} A more refined quantification of
entanglement loss can be obtained by imposing the
monotonicity of the derivative of the entanglement entropy when varying
parameter ``$t$''. That is, the condition

\begin{equation}
\frac{{\rm d}S}{{\rm d}t} < 0
\label{monotomic}
\end{equation}
implies a stronger condition in the structure of the ground state under
deformations of the parameter. This monotonic behavior of the
entanglement entropy is what we shall call monotonous entanglement loss. 

\bigskip

\paragraph{Fine-grained entanglement loss.-} When monotonous entanglement loss
holds, we can wonder whether, in fact, it is the spectra of
the underlying reduced density matrix the one that becomes more and more ordered as we
change the value of the parameter. It is then natural to ask if it is
possible to characterize the reordering of the density matrix eigenvalues
along the flow beyond the simple entropic inequality discussed before and
thereby unveil some richer structure. The finest notion of reordering when
changing the parameter is then given by the monotonic majorization (see
appendix A) of the eigenvalue distribution along the flow. If we call
$\vec{\rho}(t)$ the vector corresponding to the probability distribution of
the spectra arising from the density operator $\rho(t)$, then the condition

\begin{equation}
\vec{\rho}(t_1) \prec \vec{\rho}(t_2) \ ,
\label{finegrained}
\end{equation}
whenever $t_2 > t_1$ will reflect the strongest possible ordering of the
ground state along the flow. This is what we call fine-grained entanglement
loss, and it is fine-grained since this condition involves a whole tower of
inequalities to be simultaneously satisfied (see appendix A). In what follows we
will see that this precise majorization condition will appear in different
circumstances when studying $(1+1)$-dimensional quantum systems.

\section{Fine-grained entanglement loss with the size of the block in $(1+1)$-dimensional CFT}

A complete analytical study of majorization relations for any
$(1+1)$-dimensional conformal field theory (without boundaries\footnote{The
  case in which boundaries are present in the system must be considered from the
  point of view of boundary conformal field theory (BCFT) \cite{bcft}.}) 
is presented in the bipartite 
scenario when the size of the considered subsystems changes, i.e., 
under deformations in the interval of the accessible 
region for one of the two parties. This size will be represented 
by the length $L$ of the space interval for which we consider the reduced
density matrix $\rho_L$ after tracing out all the degrees of freedom
corresponding to the rest of the universe. Our main result in this section can
be casted into the following theorem:

\bigskip

{\bf Theorem:} $\rho_L \prec \rho_{L'}$ if $L \ge L'$ for all possible 
$(1+1)$-dimensional \emph{CFT}. 

\bigskip

\emph{Proof:}

Let $Z(\tau, \bar{\tau}) = q^{-b} {\rm tr} (q^{(L_0 + \bar{L}_0)})$ be the
partition function of a subsystem of size $L$ on a torus \cite{entropycft}, 
where $q = e^{2 \pi i \tau}$,  
$\tau = \frac{i \kappa}{\ln{(L/\epsilon)}}$ with $\kappa$ a positive 
constant, $\epsilon$ being an ultraviolet cut-off and $b \equiv (c + \bar{c})/24$
a combination of the holomorphic and antiholomorphic central 
charges that define the universality class of the model. The unnormalized 
density matrix $\rho_L$ can then be written as $\rho_L= q^{-b} q^{(L_0 +
  \bar{L}_0)}$, since $\rho_L$ can be understood as a propagator and $(L_0 +
\bar{L}_0)$ is the generator of translations in time (dilatations in the 
conformal plane) \cite{entropycft}. Furthermore, we have that 
\be
{\rm tr}(q^{(L_0 + \bar{L}_0)}) = 1 + n_1q^{\alpha_1} + n_2q^{\alpha_2}+  \cdots \ ,  
\label{trace}
\ee
due to the fact that $(L_0 + \bar{L}_0)$ is diagonal in terms of highest-weight 
states $|h, \, \bar{h}\rangle$: $(L_0 + \bar{L}_0)  |h, \, \bar{h}\rangle = 
(h+\bar{h}) |h, \, \bar{h}\rangle$, with $h\ge 0$ and $\bar{h} \ge 0$; the 
coefficients $\alpha_1, \alpha_2, \ldots > 0$, $\alpha_{i+1} > \alpha_i \
\forall i$ are related with the scaling dimensions of the descendant 
operators, and $n_1,n_2, \ldots$ are degeneracies. The normalized 
distinct eigenvalues of $\rho_L$ are then given by
\be
\begin{split}
\lambda_1 &= \frac{1}{(1 + n_1q^{\alpha_1} + n_2q^{\alpha_2} +  \cdots)}  \\
\lambda_2 &= \frac{q^{\alpha_1}}{(1 + n_1 q^{\alpha_1} + n_2 q^{\alpha_2} + \cdots)}  \\
\vdots  \\
\lambda_l &= \frac{q^{\alpha_{(l-1)}}}{(1 + n_1q^{\alpha_1} + n_2q^{\alpha_2} + \cdots)}.
\label{ei}
\end{split}
\ee

Let us define $\tilde{Z}(q) \equiv {\rm tr}(q^{(L_0 + \bar{L}_0)}) =  
(1 + n_1q^{\alpha_1} + n_2q^{\alpha_2} + \cdots )$. The behavior of the 
eigenvalues in terms of deformations with respect to parameter $L$ follows from, 
\be
\frac{{\rm d}\tilde{Z}(q)}{{\rm d}L} =\frac{\tilde{Z}(q)-1}{q} \frac{ {\rm
    d}q}{{\rm d} \ln{(L/\epsilon)}} \frac{{\rm d} \ln{(L/\epsilon)}}{{\rm d}L}
\ge 0,
\label{zet}
\ee
and therefore
\be
\frac{ {\rm d} \lambda_1}{{\rm d}L} = \frac{{\rm d}}{{\rm d}L} \left(
\frac{1}{\tilde{Z}(q)} \right) \le 0.
\label{first}
\ee
Because $\lambda_1$ is always the biggest eigenvalue $\forall L$, the first cumulant
automatically satisfies continuous majorization when decreasing the size of the interval
$L$. The variation of the rest of the eigenvalues $\lambda_l$, $l > 1$, with respect to
$L$ reads as follows:
\be
\begin{split}
&\frac{ {\rm d} \lambda_l}{{\rm d}L} = \frac{{\rm d}}{{\rm d}L}
\left( \frac{q^{\alpha_{(l-1)}}}{\tilde{Z}(q)} \right)  \\ 
&= \frac{q^{\alpha_{(l-1)}-1}}{\tilde{Z}(q)} \left(\alpha_{(l-1)}
- \frac{\tilde{Z}(q)-1}{\tilde{Z}(q)} \right) \frac{{\rm d}q}{{\rm d}L}.
\label{bigone}
\end{split}
\ee
Let us focus on the second eigenvalue $\lambda_2$. Clearly two different
situations can happen: 
\begin{itemize} 
\item{if $\left(\alpha_{1} - \frac{\tilde{Z}(q)-1}{\tilde{Z}(q)} \right) \ge
  0$, then since $\alpha_{(l-1)} > \alpha_1 \ \forall l > 2$, we have that 
 $\left(\alpha_{(l-1)} - \frac{\tilde{Z}(q)-1}{\tilde{Z}(q)} \right) > 0 \
  \forall l > 2$, which in turn implies that $\frac{ {\rm d} \lambda_l}{{\rm
  d}L} \ge 0 \ \forall l \ge 2$. From this we have that the second cumulant
  satisfies 
\be
\frac{ {\rm d} (\lambda_1+\lambda_2)}{{\rm d}L} = -\frac{{\rm d}}{{\rm d}L} \left(
 \sum_{l > 2} \lambda_l \right) \le 0 \ ,
\label{second}
\ee
thus fulfilling majorization. The same conclusion extends easily in this case to all
the remaining cumulants, and therefore majorization is satisfied by the whole
probability distribution.}
\item{if $\left(\alpha_{1} - \frac{\tilde{Z}(q)-1}{\tilde{Z}(q)} \right) <
  0$, then $\frac{ {\rm d} \lambda_2}{{\rm d}L} < 0$, and therefore  
$\frac{ {\rm d} (\lambda_1+\lambda_2)}{{\rm d}L} < 0$, so the second cumulant
  satisfies majorization, but nothing can be said from this about the rest of the
  remaining cumulants.}
\end{itemize}
Proceeding with this analysis for each one of the eigenvalues we see that, if these are
monotonically decreasing functions of $L$ then majorization is fulfilled for
the particular cumulant under study, but since $\alpha_{i+1} > \alpha_i \
\forall i$ we notice that once the first monotonically increasing eigenvalue is found,
majorization is directly satisfied by the whole distribution of eigenvalues,
therefore $\rho_L \prec \rho_{L'}$ if $L \ge L' $. This proof is 
valid for all possible $(1+1)$-dimensional conformal field theories since 
it is based only on completely general assumptions. $\square$

\subsection{Analytical finite-$L$ majorization for the critical quantum
  $XX$-model with a boundary}

Let us give an example of a similar situation to the one presented in the
previous theorem for the case of the quantum $XX$-model with a boundary, 
for which the exact spectrum of $\rho_L$ can be explicitly
computed. The Hamiltonian of the model without magnetic field, is given 
by the expression
\be
H = \sum_{n=1}^{\infty} (\sigma_n^x \sigma_{n+1}^x + \sigma_n^y
\sigma_{n+1}^y ). 
\ee
The system described by the $XX$-model is critical since it has no mass gap.
Taking the ground state and tracing 
out all but a block of $L$ contiguous spins, the density matrix $\rho_L$
describing this block can be written, in the large $L$ limit, as a thermal 
state of free fermions (see \cite{peschel1}):
\be
\rho_L = \frac{e^{-H'}}{Z_L} ,
\end{equation}
$Z_L$ being the partition function for a given $L$, $H' = \sum_{k =0}^{L-1} 
\epsilon_k d^{\dag}_k d_k$, with fermionic operators $d^{\dag}_k$, $d_k$ and 
dispersion relation
\be
\epsilon_k = \frac{\pi^2}{2 \ {\rm ln}L} (2k+1) \ \ k = 0, 1, \ldots, L-1 \ .
\ee
The eigenvalues of the density matrix $\rho_L$ can then be written in terms of 
non-interactive fermionic modes
\be
\begin{split}
\rho_L(n_0, n_1, \ldots, n_{L-1}) &= \frac{1}{Z_L} e^{-\sum_{k = 0}^{L-1} n_k \epsilon_k} \\
&= \rho_L(n_0)  \cdots \rho_L(n_{L-1}) \ ,
\end{split}
\ee
with $\rho(n_\alpha) = \frac{1}{Z_L^{\alpha}}e^{-n_{\alpha}
  \epsilon_{\alpha}}$, where $Z_L^{\alpha} = (1 + e^{-\epsilon_{\alpha}})$ is 
the partition function for the mode $\alpha$, and $n_{\alpha} = 0, 1$, $\forall \alpha$. It is worth 
noticing that the partition function of the whole block $Z_L$ can then be 
written as a product over the $L$ modes:
\begin{equation}
Z_L = \prod_{k = 0}^{L-1} \left( 1 + e^{-\epsilon_k} \right) \ .
\end{equation}

Once the density matrix of the subsystem is well characterized with
respect to its size $L$, it is not difficult to prove that $\rho_L \prec \rho_{L'}$ 
if $L \ge L'$. In order to see this, we will fix the attention in the
majorization within each mode and then we will apply the direct product lemma
from appendix A for the whole subsystem. We initially have to 
observe the behavior in $L$ of the biggest probability defined by each individual 
distribution for each one of the modes, that is, $P^{\alpha}_L =
1/Z^{\alpha}_L = (1 + e^{-\epsilon_{\alpha}})^{-1}$, for $\alpha= 0, 1, 
\ldots, L-1$. It is straightforward to see that 
\begin{equation}
\frac{{\rm d} P^{\alpha}_L}{{\rm d}L} = \frac{e^{-\epsilon_{\alpha}}}{ \left(1 + 
e^{-\epsilon_{\alpha}}\right)^2} \frac{{\rm d} \epsilon_{\alpha}}{{\rm d} L} < 0 \ ,
\end{equation}
which implies that $P^{\alpha}_L$ decreases if $L$ increases $\forall
\alpha$. This involves majorization within each mode $\alpha = 0, 1, 
\ldots, L-2$ when decreasing $L$ by one unit. In addition, we need to 
see what happens with the last mode $\alpha = L-1$ when 
the size of the system is reduced from $L$ to $L-1$. Because this mode 
disappears for the system of size $L-1$, its probability distribution 
turns out to be represented by the probability vector $(1,0)$, which 
majorizes any probability distribution of two components. Combining 
these results with the direct product lemma from appendix A, we conclude 
that this example for the quantum \emph{XX}-model provides a similar situation
for a model with a boundary to the one presented in our previous theorem.

\section{Fine-grained entanglement loss along uniparametric flows in
  $(1+1)$-dimensional quantum systems}

We study in this section strict continuous majorization relations along 
uniparametric flows, under the conditions of integrable deformations 
and monotonicity of the eigenvalues in parameter space. The main result of
this section can be casted into the next theorem: 

\bigskip

{\bf Theorem:}
consider a $(1+1)$-dimensional physical theory which depends on a set of real 
parameters $\vec{g} = (g_1,g_2,\ldots)$, such that
\begin{itemize}
\item there is a non-trivial conformal point $\vec{g}^*$, for which the model 
is conformally invariant
\item the deformations from $\vec{g}^*$ in parameter space in the positive 
direction of a given unitary vector $\hat{e}$ preserve part of the conformal 
structure of the model, i.e., the eigenvalues of the reduced density  
matrices of the vacuum $\rho(\vec{g_2})$ are still of the form given in 
eq.(\ref{ei}) for values of the parameters $\vec{g}_1 = \vec{g}^* + a \hat{e}$ 
\item $\hat{e} \cdot \left( \vec{\nabla}_{\vec{g}} q(\vec{g})\right) 
\bigg|_{\vec{g_1}} \le 0$, where $q(\vec{g})$ are the corresponding 
parameter-dependent conformal $q$-factors.
\end{itemize}

Then, away from the conformal point there is continuous majorization of the eigenvalues 
of the reduced density matrices of the ground state along the flow in 
the parameters $\vec{g}$ in the positive direction of $\hat{e}$, i.e.,
\be
\begin{split}
&\rho(\vec{g}_1) \prec \rho(\vec{g}_2), \\
\vec{g}_1= \vec{g}^* + a \hat{e}&, \, \vec{g}_2 = \vec{g}^* + a' \hat{e}, \, a' \ge a.
\end{split}
\ee

\bigskip

\emph{Proof.-}

If the eigenvalues are assumed to be of the form given by eq.(\ref{ei}), 
then it is straightforward to see that $\hat{e} \cdot \left(
\vec{\nabla}_{\vec{g}} \lambda_1(\vec{g})\right) \bigg|_{\vec{g_1}} \ge 0$, 
which assures that the first cumulant fulfills majorization. The rest of
the analysis is completely equivalent to the one presented in the previous
proof of the theorem in sec.III, which also proves this theorem. $\square$

The applicability of this theorem is based on the conditions we had to 
assume as hypothesis. Indeed, these conditions are naturally fulfilled 
by many interesting models. We now wish to illustrate this point with 
the analytical examples of similar situations for the \emph{Heisenberg} 
and  \emph{XY} quantum spin chains with a boundary.
 
\subsection{Analytical majorization along the anisotropy flow for the
  Heisenberg quantum spin chain with a boundary}

Consider the Hamiltonian of the Heisenberg quantum spin chain with a boundary
\begin{equation}
H = \sum_{n=1}^{\infty} \left(\sigma_n^x \sigma_{n+1}^x +  \sigma_n^y \sigma_{n+1}^y
+ \Delta \sigma_n^z \sigma_{n+1}^z \right) \ , 
\end{equation}
where $\Delta$ is the anisotropy parameter. This model is non-critical 
for $\Delta > 1$ and critical at $\Delta = 1$. From the pure ground state of 
the system, it is traced out half of it, getting an infinite-dimensional 
density matrix which describes half of the system ($N/2$ contiguous spins 
in the limit $N \rightarrow \infty$). The resulting reduced density 
matrix $\rho_{\Delta}$ can be written as a thermal density matrix of 
free fermions \cite{peschel1}, in such a way that its eigenvalues are given by 
\be
\rho_{\Delta}(n_0, n_1, \ldots, n_{\infty}) = \frac{1}{Z_{\Delta}} e^{-
    \sum_{k = 0}^{\infty} n_{k} \epsilon_{k}} \ ,
\ee
with dispersion relation
\be
\epsilon_k = 2 k \ {\rm arcosh} (\Delta) \ ,
\ee
and $n_k = 0,1$, for $k = 0, 1, \ldots, \infty$. The physical branch of the function ${\rm
  arcosh}(\Delta)$ is defined for $\Delta \ge 1$ and is a monotonic increasing 
function as $\Delta$ increases. On top, the whole partition function $Z_{\Delta}$ can 
be decomposed as an infinite direct product of the different free fermionic modes. 

From the last equations, it is not difficult to see that $\rho_{\Delta} 
\prec \rho_{\Delta'}$ if $\Delta \le \Delta'$. Fixing the attention in a 
particular mode $\alpha$, we evaluate the derivative of the biggest
probability for this mode, $P^{\alpha}_{\Delta} = (1 + e^{-
  \epsilon_{\alpha}})^{-1}$. This derivative is seen to be
\be
\frac{{\rm d} P^{\alpha}_{\Delta}}{{\rm d}\Delta} = \frac{2 \alpha}{(1 +
  e^{-\epsilon_{\alpha}})^2 \sqrt{\Delta^2 - 1}}> 0 \ ,
\ee
for $\alpha = 1, 2, \ldots \infty$ and $0$ for $\alpha = 0$. It follows 
from this fact that all the modes independently majorize their respective 
probability distributions as $\Delta$ increases, with the peculiarity that 
the $0$th mode remains unchanged along the flow, being its probability 
distribution always $(\frac{1}{2},\frac{1}{2})$. The particular behavior 
of this mode is the responsible for the appearance of the ``cat" state 
that is the ground state for large values of $\Delta$ (in that limit, 
the model corresponds to the quantum Ising model without magnetic field). These results, 
together with the direct product lemma from appendix A, make this example obey
majorization along the flow in the parameter.

\subsection{Analytical majorization along uniparametric flows for the quantum
  $XY$-model with a boundary}

\begin{figure}[h]
\centering
\includegraphics[width=.5\textwidth]{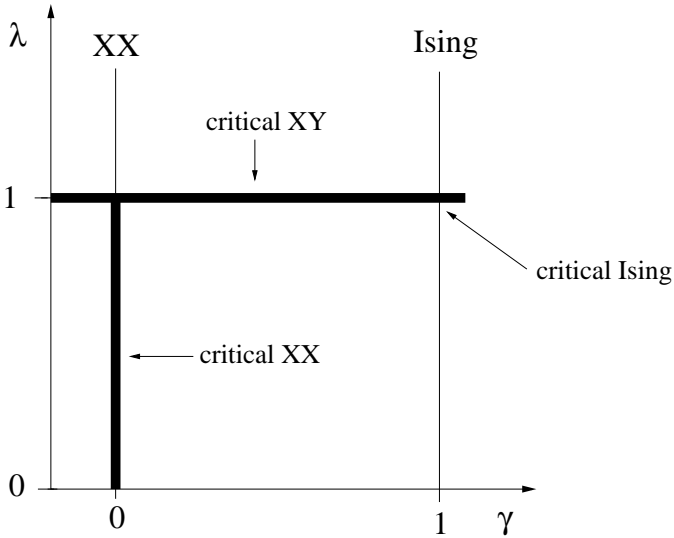}
\caption{phase diagram of the quantum $XY$-model.}
\label{phaseXY}
\end{figure}

Similar results to the one obtained for the Heisenberg model can be obtained as 
well for a more generic quantum spin chain.  Let us consider the quantum 
$XY$-model with a boundary, as described by the Hamiltonian
\begin{equation}
H = - \sum_{n = 1}^{\infty} \left( (1+\gamma) \sigma_n^x \sigma_{n+1}^x + 
(1-\gamma) \sigma_n^y \sigma_{n+1}^y
+ 2\lambda \sigma_n^z \right) \ ,
\end{equation}
where $\gamma$ can be regarded as the anisotropy parameter and $\lambda$ as 
the magnetic field. The phase diagram of this model is shown in
fig.(\ref{phaseXY}), where it is seen that there exist different critical
regions depending on the values of the parameters. Consider the ground state 
of this Hamiltonian of infinite number of spins, and trace out half of the
system (if the size of the system is $N$, we trace out $N/2$ contiguous
spins, and take the limit $N \rightarrow \infty$), for given values of 
$\lambda$ and $\gamma$. The resulting density matrix $\rho_{(\lambda,\gamma)}$ 
can be written as a thermal state of free fermions, and its eigenvalues are 
given by (see \cite{peschel1}):
\be
\rho_{(\lambda,\gamma)}(n_0, n_1, \ldots, n_{\infty}) =
\frac{1}{Z_{(\lambda,\gamma)}} 
e^{- \sum_{k = 0}^{\infty} n_{k} \epsilon_{k}} \ ,
\ee
where $n_{k} = 0,1$, and the single-mode energies $\epsilon_{k}$ are given by 
\be
\epsilon_{k} = 
\begin{cases}
2 k  \epsilon \ , & {\rm if} \ \lambda < 1 \\
(2 k+1)  \epsilon \ , & {\rm if} \ \lambda > 1 \ , 
\end{cases}
\label{dispersion}
\ee
with $k = 0, 1, \ldots , \infty$. The parameter $\epsilon$ is defined by the relation
\be
\epsilon = \pi \frac{I(\sqrt{1-x^2})}{I(x)} \ ,
\ee
$I(x)$ being the complete elliptic integral of the first kind
\be
I(x) = \int_0^{\pi/2} \frac{d \theta}{\sqrt{1 - x^2 \sin^2 (\theta)}}
\ee
and $x$ being given by 
\be
x = 
\begin{cases}
(\sqrt{\lambda^2 + \gamma^2 - 1})/\gamma \ , & {\rm if} \ \lambda < 1 \\
\gamma / (\sqrt{\lambda^2 + \gamma^2 - 1})  \ , & {\rm if} \ \lambda > 1 \ , 
\end{cases}
\label{eks}
\ee
with the condition $\lambda^2 + \gamma^2 > 1$ (external region of the BM-circle
\cite{BM}). 

We note that the probability distribution defined by the eigenvalues of 
$\rho_{(\lambda, \gamma)}$ is the direct product of distributions for each 
one of the separate modes. Therefore, in order to study majorization we can 
focus separately on each one of these modes, in the same way as we already did
in the previous 
examples. We wish now to consider our analysis in terms of the flows 
with respect to the magnetic field $\lambda$ and with respect to the
anisotropy $\gamma$ in a separate way. 

\subsubsection{Flow along the magnetic field $\lambda$}

We consider in this subsection a fixed value of $\gamma$ while the value of 
$\lambda$ changes, always fulfilling the condition $\lambda^2+ \gamma^2 >
1$. Therefore, at this point we can drop $\gamma$ from our notation.
We separate the analysis of majorization for the regions $1 < \lambda <
\infty$ and $+\sqrt{1-\gamma^2} < \lambda < 1$ for reasons that will become 
clearer during the example but that already can be realized just by looking at 
the phase space structure in fig.(\ref{phaseXY}).

\bigskip 

\paragraph{$1 < \lambda < \infty$.-} We show that $\rho_{\lambda} \prec
\rho_{\lambda'}$ if $\lambda \le \lambda'$. In this region of parameter 
space, the biggest probability for the mode $\alpha$ is $P^{\alpha}_{\lambda} 
= (1+e^{- \epsilon_{\alpha}})^{-1}$, with
\be
\epsilon_{\alpha} = (2 \alpha + 1) \pi \frac{I(\sqrt{1-x^2})}{I(x)}  = 
(2 \alpha + 1) \epsilon \ ,
\ee
where $x = \gamma/(\sqrt{\lambda^2 + \gamma^2 - 1})$. The variation of the 
biggest eigenvalue with respect to $\lambda$ is
\be
\frac{{\rm d}P^{\alpha}_{\lambda}}{{\rm d} \lambda} =  \frac{ (2\alpha + 1) e^{-(2
\alpha+1) \epsilon}}{\left( 1 + e^{- (2 \alpha + 1) \epsilon} \right)^2} 
\frac{{\rm d} \epsilon}{{\rm d}\lambda} \ .
\ee
It is easy to see that 
\be
\begin{split}
&\frac{{\rm d} \epsilon}{{\rm d}\lambda} =  \frac{{\rm d} \epsilon}{{\rm d}x}
  \frac{{\rm d} x}{{\rm d}\lambda} \\
&= \frac{\pi}{I(x)} \left(\frac{{\rm d} I (\sqrt{1-x^2})}{{\rm d} x} - 
\left(\frac{I(\sqrt{1-x^2})}{I(x)}\right) \frac{{\rm d} I(x)}{{\rm d}x}
  \right) \frac{{\rm d}x}{{\rm d}\lambda} > 0 \ , 
\end{split}
\ee
since both $\left( \frac{{\rm d} \epsilon}{{\rm d}x} \right) < 0$ and
$\left(\frac{{\rm d}x}{{\rm d}\lambda}\right) < 0$. 
Therefore, $\frac{{\rm d}P^{\alpha}_{\lambda}}{{\rm d} \lambda} > 0$ for $\alpha = 0,1,
\ldots, \infty$. This derivation shows mode-by-mode majorization when
$\lambda$ increases. Combining this result with the direct product 
lemma from appendix A, we see that this example obeys majorization.

\bigskip

\paragraph{ $+\sqrt{1-\gamma^2} < \lambda < 1$.-} For this case, we show 
that $\rho_{\lambda} \prec \rho_{\lambda'}$ if $\lambda \ge \lambda'$. 
In particular, the probability distribution for the $0$th fermionic mode 
remains constant and equal to $(\frac{1}{2},\frac{1}{2})$, which brings
a ``cat'' state for low values of $\lambda$. Similar to the latter case, 
the biggest probability for mode $\alpha$ 
is $P^{\alpha}_{\lambda} = (1+e^{-\epsilon_{\alpha}})^{-1}$, with
\be
\epsilon_{\alpha} = 2 \alpha \pi \frac{I(\sqrt{1-x^{2}})}{I(x)} = 2\alpha \epsilon \ ,
\ee
and $x = (\sqrt{\lambda^2 + \gamma^2 - 1})/\gamma$. 
Its derivative with respect to $\lambda$ is
\be
\frac{{\rm d}P^{\alpha}_{\lambda}}{{\rm d} \lambda} = \frac{ 2\alpha  e^{-2 \alpha
\epsilon}}{\left( 1 + e^{- 2 \alpha \epsilon} \right)^2} \frac{{\rm
    d}\epsilon}{{\rm d}\lambda} \ .
\label{deri}
\ee

It is easy to see that this time $\left(\frac{{\rm d}\epsilon}{{\rm d}\lambda}\right) <
0$, and therefore $\frac{{\rm d}P^{\alpha}_{\lambda}}{{\rm d} \lambda} < 0$ for $\alpha =
1, 2, \ldots, \infty$, which brings majorization individually for each one of 
these modes when $\lambda$ decreases. The mode $\alpha = 0$ needs of special 
attention, from eq.(\ref{deri}) it is seen that
$\frac{{\rm d}P^{\alpha=0}_{\lambda}}{{\rm d} \lambda} = 0$, therefore the probability 
distribution for this mode remains constant and equal to $(\frac{1}{2},
\frac{1}{2})$ all along the flow. This is a marginal mode that brings the 
system to a ``cat" state that appears as ground state of the system 
for low values of $\lambda$. Notice that this peculiarity is rooted on the 
particular form of the dispersion relation given in equation
(\ref{dispersion}), which is proportional to $2 k$ instead of $2k + 1$ 
for this region in parameter space. These results, together with the 
direct product lemma from appendix A, prove that this example fulfills also
majorization. 

\subsubsection{Flow along the anisotropy $\gamma$}

In this subsection, the magnetic field $\lambda$ is fixed and the 
anisotropy $\gamma$ is the only free parameter of the model, always 
fulfilling $\lambda^2 + \gamma^2 > 1$. Thus, at this point we can drop
$\lambda$ from our notation. We will see that $\rho_{\gamma} 
\prec \rho_{\gamma'}$ if $\gamma \ge \gamma'$, in the two regions $1 < 
\lambda < \infty$ and $+\sqrt{1-\gamma^2} < \lambda < 1$. In particular, 
in the region  $+\sqrt{1-\gamma^2} < \lambda < 1$, the probability
distribution for the $0$th fermionic mode remains constant and equal 
to $(\frac{1}{2},\frac{1}{2})$. Let us consider the biggest probability 
for the mode $\alpha$,
$P^{\alpha}_{\gamma} =  (1+e^{- \epsilon_{\alpha}})^{-1}$, with 
$\epsilon_{\alpha} = \omega \epsilon$, where 
\be
\omega = 
\begin{cases}
2 \alpha \ , & {\rm if} \ \lambda < 1 \\
(2 \alpha + 1)  \ , & {\rm if} \ \lambda > 1 \ , 
\end{cases}
\label{omega}
\ee
and $\epsilon$ as defined in the preceding sections. It is easy to verify that 
\be
\frac{{\rm d} P^{\alpha}_{\gamma}}{{\rm d} \gamma} =  \frac{\omega e^{- \omega
    \epsilon_{\alpha}}}{(1+e^{-\omega \epsilon_{\alpha}})^2} \frac{{\rm d}
  \epsilon}{{\rm d} x} \frac{{\rm d} x}{{\rm d} \gamma}  < 0  
\label{proba}
\ee
for $\alpha = 0, 1, \ldots, \infty$ if $\lambda > 1$ and for $\alpha = 1, 2,
\ldots, \infty$ if $\lambda < 1$. The mode $\alpha = 0$ for $\lambda < 1$
needs of special attention: it is seen that $\frac{{\rm
    d}P^{\alpha=0}_{\lambda}}{{\rm d}
  \lambda} = 0$, therefore the probability distribution for this mode remains 
constant and equal to $(\frac{1}{2},\frac{1}{2})$ all along the flow. 
These results, together with the direct product lemma from appendix A, 
show that this case also obeys majorization along the flow in the parameter.

\section{Conclusions}

In this paper we have provided in a rigorous way
explicit proves for all the majorization 
conjectures raised by Latorre, L$\ddot{{\rm u}}$tken, Rico, Vidal and 
Kitaev in previous papers on quantum spin chains \cite{fine-grained,
  VLRK02}. In particular, we have developed 
a completely general proof of
majorization relations underlying the structure of the vacuum with respect 
to the size of the block $L$ for all possible
$(1+1)$-dimensional conformal field theories. An example of a similar
situation has been given with the particular case of the \emph{XX}-model with
a boundary, for which the explicit calculation of the
eigenvalues of the reduced density matrix can be performed. We have proven as
well the existence of a fine-grained entanglement loss for $(1+1)$-dimensional 
quantum systems along uniparametric flows, regarded that
perturbations in parameter space preserve part of the
conformal structure of the partition function, and some monotonicity
conditions hold as well. Again examples of similar situations have been
provided by means of the Heisenberg
and \emph{XY} models with a boundary. Our results provide solid mathematical grounds for the
existence of majorization relations along RG-flows underlying the structure of
the vacuum of (1+1)-dimensional quantum spin chains. 

Understanding the entanglement structure of the vacuum of $(1+1)$-dimensional models is
a major task in quantum information science. For instance, spin chains like
the ones described in the particular examples of this paper can be used as
possible approximations to the complicated interactions that take place in the register of
a quantum computer \cite{porras}. Entanglement across a quantum phase transition has also an
important role in quantum algorithm design, and in particular in quantum
algorithms by adiabatic evolution \cite{orus1}. On top, the properties of quantum state
transmission through spin chains are also intimately related to the
entanglement properties present in the chain \cite{tobby}. Consequently, our precise characterization of
entanglement in terms of majorization relations should be helpful for the
design of more powerful quantum algorithms and quantum state transmission protocols. 

It would also be of interest
trying to relate the results presented in this paper to possible
extensions of the c-theorem \cite{ctheorem} to systems with more than
(1+1)-dimensions. While other approaches are also possible \cite{ignacio-forte}, 
majorization may be a unique tool in
order to envisage irreversibility of RG-flows in terms of 
properties of the vacuum only, and
some numerical results in this direction have already been observed in
systems of different dimensionality along uniparametric flows \cite{orus}. 
New strict mathematical results could probably be achieved in these
situations following the ideas that we have presented all along this work.   

\vspace{5pt}

{\bf Acknowledgments:} the author is grateful to very fruitful and
enlightening discussions with
J.I. Latorre, C. A. L$\ddot{{\rm u}}$tken, E. Rico and G. Vidal 
about the content of this paper, and also to H. Q. Zhou, T. Barthel, 
J. O. Fjaerestad and U. Schollwoeck for pointing an error in a previous
version of this paper.  Financial
support from projects MCYT FPA2001-3598, GC2001SGR-00065 and IST-199-11053 is
also acknowledged.

\vspace{5pt}

{\bf Note added:} after completion of this paper a similar work appeared
\cite{bcft} in which entanglement and majorization are considered from 
the point of view of boundary conformal field theory, and where it was noticed
that there was an error in the proof of the first theorem of a previous 
version of this paper. That error has been corrected in this new version.

\appendix

\section{Lemmas on Majorization}

This appendix includes the formal definitions of majorization  \cite{maj} 
as well as a lemma that is used along the examples presented in this work.

\subsection{Definitions}

Let $\vec{x}$, $\vec{y}\in \mathbb{R}^N$ be two vectors such that 
$\sum_{i = 1}^N x_i = \sum_{i=1}^N y_i = 1$, which represent two 
different probability distributions. We say that distribution $\vec{y}$ 
majorizes distribution $\vec{x}$, written $\vec{x}\prec \vec{y}$, 
if and only if there exist a set of permutation matrices $\{P_j\}$ 
and probabilities $p_j \ge 0$, $\sum p_j=1$, such that \begin{equation}
\vec{x} = \sum_j p_j P_j \vec{y} \ .
\label{defone}
\end{equation}
Since, from the previous definition, $\vec{x}$ can be obtained by means 
of a probabilistic combination of permutations of $\vec{y}$, we get the
intuitive notion that distribution $\vec{x}$ is 
more disordered than $\vec{y}$.

Notice that in (\ref{defone}), $\sum_j p_j P_j=D$ defines a doubly 
stochastic matrix, i.e. $D$ has nonnegative entries and each row and 
column sums to unity. Then,  $\vec{x}\prec \vec{y}$  if and only if 
$\vec{x} = D \vec{y}$, $D$ being a doubly stochastic matrix. 

Another equivalent definition of majorization can be stated in terms 
of a set of inequalities between the two distributions. Consider the 
components of the two vectors sorted in decreasing order, written as 
$(z_1, \ldots z_N) \equiv \vec{z}^\downarrow$. Then, $\vec{x}^\downarrow 
\prec \vec{y}^\downarrow$ if and only if

\begin{equation}
\sum_{i=1}^k x_i \leq \sum_{i=1}^k y_i \qquad k = 1 \ldots N \ .
\label{deftwo}
\end{equation}
All along this work, these probability sums are called cumulants.

A powerful relation between majorization and any convex function $f$ over the
set of probability vectors 
states that $\vec{x} \prec \vec{y} \Rightarrow f \left(\vec{x}\right) 
\le f\left(\vec{y}\right)$. From this relation it follows that the 
common Shannon entropy $H(\vec{x}) \equiv -\sum_{i=1}^N x_i \log{x_i}$ 
of a probability distribution satisfies $H \left(\vec{x}\right) 
\ge H\left(\vec{y}\right)$ whenever $\vec{x} \prec \vec{y}$. In what follows 
we present a lemma that is used all along our work in the different examples
that we analyze.

\subsection{Direct product lemma \cite{fine-grained}}

If $\vec{p}_1 \prec \vec{p}_2$, $\vec{q}_1 \prec \vec{q}_2$ then 
$(\vec{p}_1 \otimes \vec{q}_1) \prec (\vec{p}_2 \otimes \vec{q}_2)$.  
This means that majorization is preserved under the direct product 
operation.

\vspace{10pt}

\emph{Proof.-}

If $\vec{p}_1 \prec \vec{p}_2$ and $\vec{q}_1 \prec \vec{q}_2$ then 
$\vec{p}_1 = D_{p} \vec{p}_2$ and $\vec{q}_1 = D_{q} \vec{q}_2$ where 
$D_{p},  D_{q}$ are both doubly stochastic matrices. Therefore 
$(\vec{p}_1 \otimes \vec{q}_1) = (D_{p} \otimes D_{q}) (\vec{p}_2 \otimes
\vec{q}_2)$, where $(D_{p} \otimes D_{q})$ is a doubly stochastic matrix 
in the direct product space, and so $(\vec{p}_1 \otimes \vec{q}_1) \prec 
(\vec{p}_2 \otimes \vec{q}_2)$. $\square$

\end{document}